\documentclass[11pt]{article}
\usepackage{verbatim}
\usepackage{amsfonts}
\usepackage{graphics}
\usepackage{amsmath}
\usepackage{times}
\usepackage{graphicx}
\usepackage{color}
\usepackage{enumerate}
\usepackage{epsfig}
\def\SS{\scriptsize}
\def\TM{\!\!$^{\mbox{\texttt{\SS TM}}}$~}

\begin{document}
\begin{center}
{\bf \large Pricing of  average strike Asian call option using numerical PDE methods\\[1cm]
Abhishek Kumar$^1$, Ashwin Waikos$^1$ and Siddhartha P. Chakrabarty$^{1}$}\\[1cm]
\end{center}
$^1$ Department of Mathematics, Indian Institute of Technology Guwahati,
Guwahati 781039, Assam, India\\
\begin{abstract}
In this paper, a standard PDE for the pricing of arithmetic average strike Asian call option is presented.
A Crank-Nicolson Implicit Method and a Higher Order Compact finite difference scheme for this 
pricing problem is derived. Both these schemes were implemented for various values of risk free rate and volatility. 
The option prices for the same set of values of risk free rate and volatility was also computed using Monte Carlo simulation.
The comparative results of the two numerical PDE methods shows close match with the Monte Carlo results, with the Higher Order Compact
scheme exhibiting a better match. To the best of our knowledge, this is the first work to use the numerical PDE approach for pricing 
Asian call options with average strike.
\end{abstract}


\section{Introduction}

\def\prob{\mathbb{P}}
\def\Fc{\mathcal{F}}
Options are one of the most common financial derivatives that are traded both in exchanges as well as over the counter 
\cite{Hull06,Wilmott95,Roman04,Capinski03}.
The two most common options are the European and the American options. 
The purchase or sale of the underlying asset for the option takes place at the 
discretion of the holder of the option for a price called the exercise price or the strike price
at a fixed time called expiration time (denoted by $T$) in case of European option and any time in $[0,T]$ in case of American option.
Options can be further classified as call or put depending whether the holder has the right to buy or the right to sell the underlying asset.
In addition there are numerous other options that can broadly be classified as exotic options \cite{Hull06,Wilmott95, Higham04,Seydel06}. The specialty
of these kinds of options is that the final payoff is more sophisticated and sometimes depends on some function of the path of prices of underlying asset.
One of the common path dependent exotic options is the Asian option \cite{Hull06,Wilmott95, Higham04,Seydel06,Shreve04}, 
whose payoff depends on the average historical stock prices. In this paper, we will focus on pricing of an Asian call option with
arithmetic average strike.

We begin with the assumption that the stock prices $S(t)$ follow a geometric Brownian motion given by the
stochastic differential equation \cite{Hull06,Wilmott95, Higham04,Seydel06,Shreve04},
\[dS(t)=rS(t)dt+\sigma S(t)dW(t),\]
where $r$ is the average rate of growth of asset prices or drift, $\sigma$ is the volatility and $W(t)(0 \le t \le T)$ is a
Brownian motion or Wiener process under the risk neutral measure $\prob$.
The payoff for an Asian {\it call option} \cite{Hull06,Wilmott95,Higham04,Seydel06,Shreve04},
with arithmetic average strike is given by, 
\[V(T)=\max\left(S(T)-\frac{1}{T}\int\limits_{0}^{T} S(u)du,0\right).\]
The price of the option at time $t\in [0,T]$ (with filtration $\Fc_{t}$) is given by
the risk-neutral pricing formula \cite{Shreve04},
\[V(t)=E\left(e^{-r(T-t)}V(T)|\Fc_{t}\right).\] 
In order to deal with the challenge of the payoff $V(T)$ being path-dependent
we introduce a stochastic process $I(t)$ \cite{Wilmott95,Shreve04} given by,
\[I(t)=\int\limits_{0}^{t} S(u)du.\]
The stochastic differential equation for evolution of $I(t)$ is thus given by,
\[dI(t)=S(t)dt.\]
Thus the Asian call option price $V(S,I,t)$ for continuous arithmetic average strike satisfies the backward PDE \cite{Wilmott95,Seydel06,Shreve04},
\[\frac{\partial V}{\partial t}+rS\frac{\partial V}{\partial S}+\frac{1}{2}\sigma^{2}S^{2}\frac{\partial^{2} V}{\partial S^{2}}
+S\frac{\partial V}{\partial I}-rV=0.\]
Note that the above problem is three dimensional which leads to greater computational costs. 
This motivates the reduction of the problem into lower dimension.
\cite{Wilmott95}.  For this purpose, a new variable 
$R(t)=\frac{1}{S(t)}\int_{0}^{t}S(u)du$ 
is defined \cite{Wilmott95,Seydel06}. 
This in turn motivates the ansatz $V(S,I,t)=S\cdot H(R,t)$ for some function $H(R,t)$. It can be shown that
the SDE satisfied by $R(t)$ is \cite{Seydel06},
\[dR(t)=\left(1+(\sigma^{2}-\mu)R(t)\right)dt-\sigma R(t)dW(t).\]
Also the function $H(R,t)$ satisfies the PDE \cite{Seydel06},
\[\frac{\partial H}{\partial t}+\frac{1}{2}\sigma^{2}R^{2}\frac{\partial^{2} H}{\partial R^{2}}+
(1-rR)\frac{\partial H}{\partial R}=0.\]
The solution of this backward PDE requires a final condition and two boundary conditions which are outlined below
\cite{Wilmott95,Seydel06}.
\begin{enumerate}[i.]
\item \textit{Final Condition :} The final payoff for the option gives the final condition,
\[H(R(T),T)=\max\left(1-\frac{1}{T}R(T),0\right).\]
\item \textit{Right Hand Boundary Condition :} The right hand boundary condition for $R\rightarrow \infty$ can be obtained
by observing that since the integral $R(t)$ is bounded, so $S\rightarrow 0$ for $R\rightarrow \infty$. For $S\rightarrow 0$ the option
is not exercised rendering it's value to be $0$. Hence, 
\[H(R,t)=0~\text{for}~R\rightarrow \infty.\]
\item \textit{Left Hand Boundary Condition :} The left hand boundary condition for $R\rightarrow 0$ can be obtained from
the similarity reduction equation. The term $R \partial H/ \partial R \rightarrow 0$ as $R\rightarrow 0$. Assuming that $H$ is
bounded it follows that the term $R^{2} \partial^{2} H/ \partial R^{2} \rightarrow 0$ as $R\rightarrow 0$. This leads to
the boundary condition,
\[\frac{\partial H}{\partial t}+\frac{\partial H}{\partial R}=0~\text{for}~R\rightarrow 0.\]
\end{enumerate}
The problem to be solved now reduces to,
\[\frac{\partial H}{\partial t}+\frac{1}{2}\sigma^{2}R^{2}\frac{\partial^{2} H}{\partial R^{2}}
+(1-rR)\frac{\partial H}{\partial R}=0.\]
subject to,
\begin{eqnarray}
H(R(T),T)&=&\max\left(1-\frac{R(T)}{T},0\right)\nonumber\\
\frac{\partial H}{\partial t}+\frac{\partial H}{\partial R}&=&0~\text{for}~R\rightarrow 0 \label{eq:model}\\
H&=&0~,\text{for}~R\rightarrow \infty. \nonumber
\end{eqnarray}
Once the solution $H(R,t)$ is obtained the price of the Asian option is determined by,
\[V(S(0),R(0),0)=S(0)H(R(0),0),\]
where $S(0)$ is the initial stock price.
For pricing of options (especially exotic options) the standard method used is Monte Carlo simulation 
\cite{Glasserman03,Boyle77,Boardie96}. This involves simulating the paths of the underlying asset and calculating the 
option price based on this path. A large number of such simulations are run and the average of the option prices 
from each simulation is taken to be the option price.
Several methodologies have been adopted for pricing of options using the numerical PDE approach.
Some of the most commonly used ones are finite differences of lower order \cite{Rogers95,Vecer01,Zvan96,Dubois05}
and higher order compact schemes for American options \cite{Tangman08,Tangman08b} and option pricing in stochastic volatility model
\cite{DuringSSRN}.

\section{Crank Nicolson Implicit Method}

The problem of pricing the average strike Asian call option essentially entails solving for equation
(\ref{eq:model}). While geometric mean Asian option admits closed form solutions \cite{Glasserman03}, the same is not
true in case of arithmetic average Asian options. As such one has to seek a solution through numerical methods for PDEs
\cite{Vecer01}. There are several articles in literature which dwell upon numerical PDE approach to Asian option pricing.
One of the first papers to deal with numerical PDE pricing of options is by Rogers and Shi \cite{Rogers95}.
In this paper, the authors first reduce the problem of solving a parabolic PDE in two variables and present a highly accurate lower bound.
Zvan et al. \cite{Zvan96} in their technical report, do an extensive analysis of numerical PDE methods of Asian options.
They discuss the shortcomings of applying the usual numerical PDE techniques used for standard options in case of Asian options.
In particular they adapt flux limiting techniques from computational fluid dynamics (CFD) to tackle the problem of spurious oscillations that
arise in Asian options. Vecer \cite{Vecer01} provided a numerical implementation of the Asian option pricing problem using the $\theta$ method.
Dubois and Lelievre \cite{Dubois05} extend the approach by Rogers and Shi \cite{Rogers95} and propose a scheme which produced
fast and accurate results. While all these papers \cite{Rogers95,Vecer01,Zvan96,Dubois05} do examine the pricing problem from the 
numerical PDE point of view, the focus is mostly on fixed strike options. Rogers and Shi \cite{Rogers95} and Zvan et al. \cite{Zvan96}
present some results on average strike put options.

Our main objective in this paper will be to use Higher Order Compact (HOC) scheme for this purpose. 
We will postpone the discussion on this until the next section. In this section we will present the Crank-Nicolson Implicit Method (CNIM)
for solving equation \eqref{eq:model} and compare the results with those obtained by Monte Carlo simulation \cite{BTP11}.

CNIM is obtained by taking the average between Forward-Time Centered-Space method (FTCS) and Backward-Time Centered-Space method (BTCS).
For this purpose, let us define a finite difference discretization of the PDE (equation(\ref{eq:model})) with the uniform grid
$t_{n}=t_{1}+(n-1)\cdot \Delta_{t},n=1:N+1$ and $R_{i}=R_{1}+(i-1)\cdot \Delta_{R},i=1:M+1$, where
$\Delta_{t}$ and $\Delta_{R}$ are the temporal and spatial mesh size respectively.
The values used in this paper are $t_{1}=0$ and $t_{N+1}=T=1$ (i.e $1$ year option) with $R_{1}=0$ and $R_{M+1}=5$.
Let us also define the variables $c(R)=\frac{1}{2}\sigma^{2}R^{2}$ and $d(R)=(1-rR)$.
$H(R,t)$ at the point $(R_{i},t_{n})$ is denoted by $H_{i}^{n}$. The CNIM discretization of 
equation (\ref{eq:model}) is then given by,
\begin{eqnarray}
\frac{H^{n+1}_{i}-H^{n}_{i}}{\Delta_{t}}&+&\frac{c_{i}}{\Delta_{R}^{2}}\left[\frac{H^{n+1}_{i+1}+H^{n}_{i+1}}{2}-2\left(\frac{H^{n+1}_{i}+ H^{n}_{i}}{2}\right)+\frac{H^{n+1}_{i-1} + H^{n}_{i-1}}{2} \right]\nonumber\\
&+&\frac{d_{i}}{2\Delta_{R}}\left[\frac{H^{n+1}_{i+1}+H^{n}_{i+1}}{2}-\frac{H^{n+1}_{i-1}+ H^{n}_{i-1}}{2}\right]=0
\label{cnim_eq}
\end{eqnarray}
The above can now be rewritten as,
\begin{equation}
G_{i}H_{i+1}^{n}+K_{i}H_{i}^{n}+J_{i}H_{i-1}^{n}=D_{i}H_{i+1}^{n+1}+E_{i}H_{i}^{n+1}+F_{i}H_{i-1}^{n+1}
\label{cnim_matrix_form_eq}
\end{equation}
where,
\begin{eqnarray*}
\begin{aligned}
G_{i}&=&-\frac{c_{i}}{2\Delta_{R}^{2}}-\frac{d_{i}}{4\Delta_{R}}\\
K_{i}&=&\frac{c_{i}}{\Delta_{R}^{2}}+\frac{1}{\Delta_{t}}\\
J_{i}&=&-\frac{c_{i}}{2\Delta_{R}^{2}}+\frac{d_{i}}{4\Delta_{R}}
\end{aligned} \qquad \qquad
\begin{aligned}
D_{i}&=&\frac{c_{i}}{2\Delta_{R}^{2}}+\frac{d_{i}}{4\Delta_{R}}\\
E_{i}&=&-\frac{c_{i}}{\Delta_{R}^{2}}+\frac{1}{\Delta_{t}}\\
F_{i}&=&\frac{c_{i}}{2\Delta_{R}^{2}}-\frac{d_{i}}{4\Delta_{R}}.
\end{aligned}
\end{eqnarray*}
Let us define the vector,
\[H^{(n)}=(H_{2}^{n},H_{3}^{n},H_{4}^{n},H_{5}^{n},\dots,H_{M}^{n})^{\top}~\text{for}~n=1:N+1.\]
The CNIM can now be written in the matrix form as,
\begin{equation}
BH^{(n)}=AH^{(n+1)}+b^{(n)},
\label{cnim_problem}
\end{equation}
where,
\[
A=\begin{pmatrix}
E_{2} & D_{2} & 0 & \dots & 0\\
F_{3} & E_{3} & D_{3} & \ddots & \vdots\\
0& \ddots & \ddots & \ddots & 0\\
\vdots & \ddots & \ddots & \ddots & D_{M-1}\\
0 & \dots & 0 & F_{M} & E_{M}
\end{pmatrix}.
\]
and\
\[
B=\begin{pmatrix}
K_{2} & G_{2} & 0 & \dots & 0\\
J_{3} & K_{3} & G_{3} & \ddots & \vdots\\
0& \ddots & \ddots & \ddots & 0\\
\vdots & \ddots & \ddots & \ddots & G_{M-1}\\
0 & \dots & 0 & J_{M} & K_{M}
\end{pmatrix}.
\]
\[
b^{(n)}=\begin{pmatrix}
F_{2} H_{1}^{n+1}-J_{2} H_{1}^{n}\\
\dots\\
\dots\\
D_{M} H_{M+1}^{n+1}-G_{M} H_{M+1}^{n}\\
\end{pmatrix}.
\]
From second order finite differences, the one-sided difference is given by,
\[\frac{\partial H}{\partial R}\biggr|_{i}=\frac{-H_{i+2}+4H_{i+1}-3H_{i}}{2\Delta_{R}}+O(\Delta_{R}^{2}).\]
Therefore, applying the same on the left boundary along with backward time approximation we get,
\[H^{n}_{1}=(1-3k)H^{n+1}_{1}+4kH^{n+1}_{2}-kH^{n+1}_{3},\]
where $k=\frac{\Delta_{t}}{2\Delta_{R}}$. The right boundary $H^{n}_{M+1}=0$ follows from equation ({\ref{eq:model}}).
The final condition is given by,
\[H_{i}^{N+1}=\left(1-\frac{R_{i}}{T}\right)^{+}.\]
The CNIM formulation above was solved using MatLab \TM. The solution was obtained using an iterative process 
which involved a tolerance criterion of $|H_{\text{new}}(1,1)-H_{\text{old}}(1,1)|<\epsilon$.
For the purpose of this implementation the number of time and space grid points were taken to be $101$ and $501$ respectively.
The tolerance level was taken to be $\epsilon=10^{-8}$.
\begin{table}[h!]
\begin{center}
\begin{tabular}{|l|l|l|l|l|l|l|}
\hline
\textbf{r $\rightarrow$} & $0.06$ & $0.06$ & $0.1$ & $0.1$ & $0.2$ & $0.2$ \\
\hline
\textbf{$\sigma\downarrow$} & \textbf{CNIM} & \textbf{MC} & \textbf{CNIM} &
\textbf{MC} &
\textbf{CNIM} & \textbf{MC} \\
\hline
0.05 & 3.5025	& 3.1509 & 5.1148	& 4.8734 & 9.3988	& 9.3486\\
& (0.375339) & (4.511535) & (0.365693)	& (4.492267) & (0.344431) &	(4.556745)\\
\hline
0.1 & 4.1353 &	4.0124 &	5.5629 &	5.4183	& 9.5333 &	9.433\\
& (0.117604) & (4.505087) & (0.11831) & (4.489719) & (0.117648) & (4.53052)\\
\hline
0.2 & 6.1337 &	6.1172 &	7.2951 & 7.2625	& 10.547 & 10.4894\\
& (0.117679) &	(4.490006) &	(0.117177)	& (4.481982) &	(0.117271) &	(4.530701)\\
\hline
0.3 & 8.3256 & 8.3155	& 9.3669	& 9.3484 &	12.2035 &	12.163\\
& (0.11757)	& (4.490814)	& (0.117839)	& (4.518666)	& (0.118941)	& (4.500043)\\
\hline
0.4 & 10.5403	& 10.5358	& 11.5081	& 11.4952	& 14.0885	& 14.0581\\
& (0.12389) &	(4.483286) &	(0.117663) &	(4.483686)	& (0.116222)	& (4.490074)\\
\hline
\end{tabular}
\end{center}
\caption{Comparison between Monte Carlo Simulation (MC) and Crank Nicolson Implicit
Method (CNIM). The values in braces represent the CPU time in seconds. The initial stock price was $S(0)=100$.
\label{MC_CNIM_Table}}
\end{table}

\begin{figure}[hb]
\centering
\includegraphics[width=0.7\textwidth]{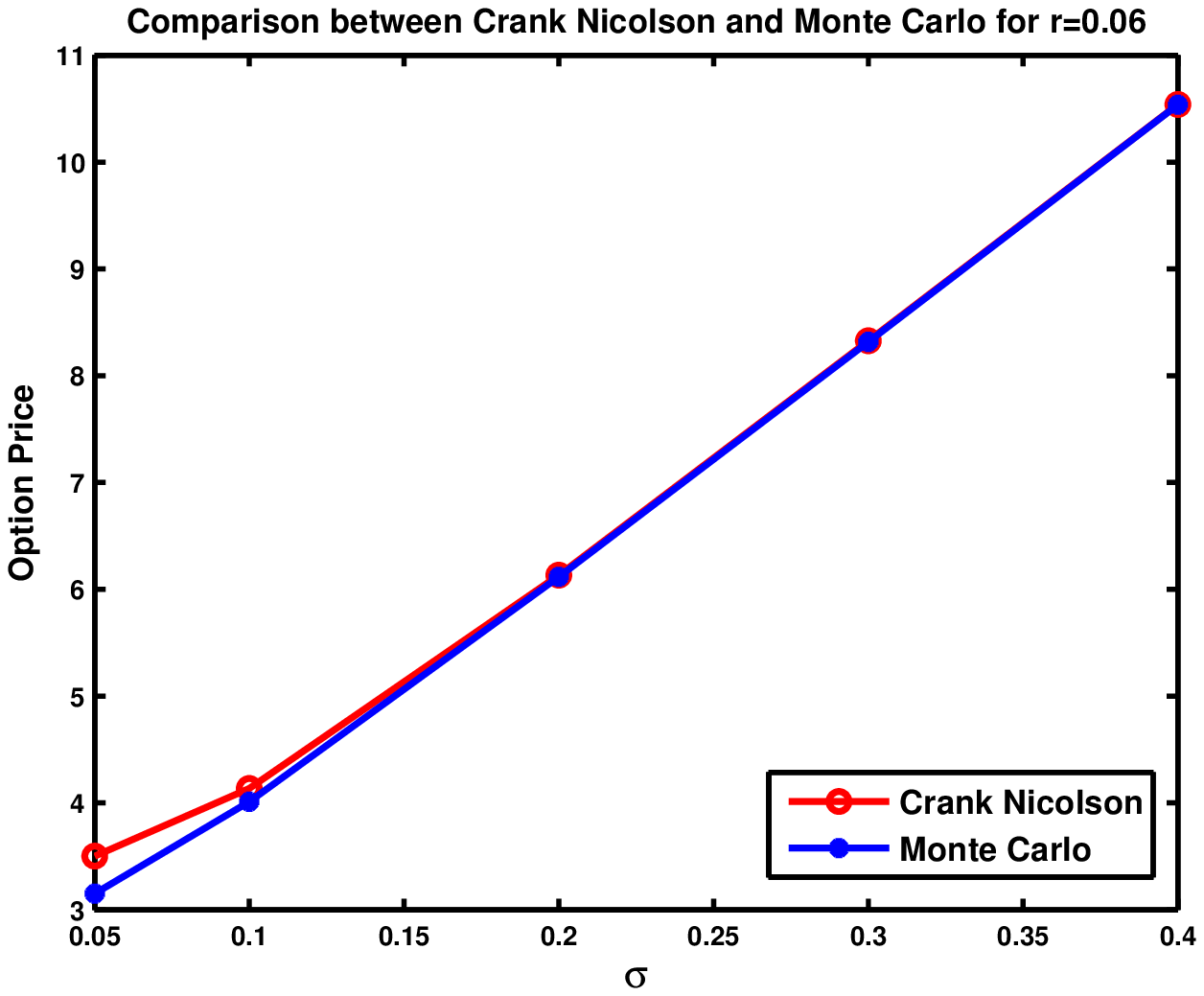}
\caption{Comparison between Monte Carlo Simulation (MC) and Crank Nicolson Implicit Method (CNIM)\label{MC_CNIM_Graph1} for r=0.06}
\end{figure}

\begin{figure}[hb]
\centering
\includegraphics[width=0.7\textwidth]{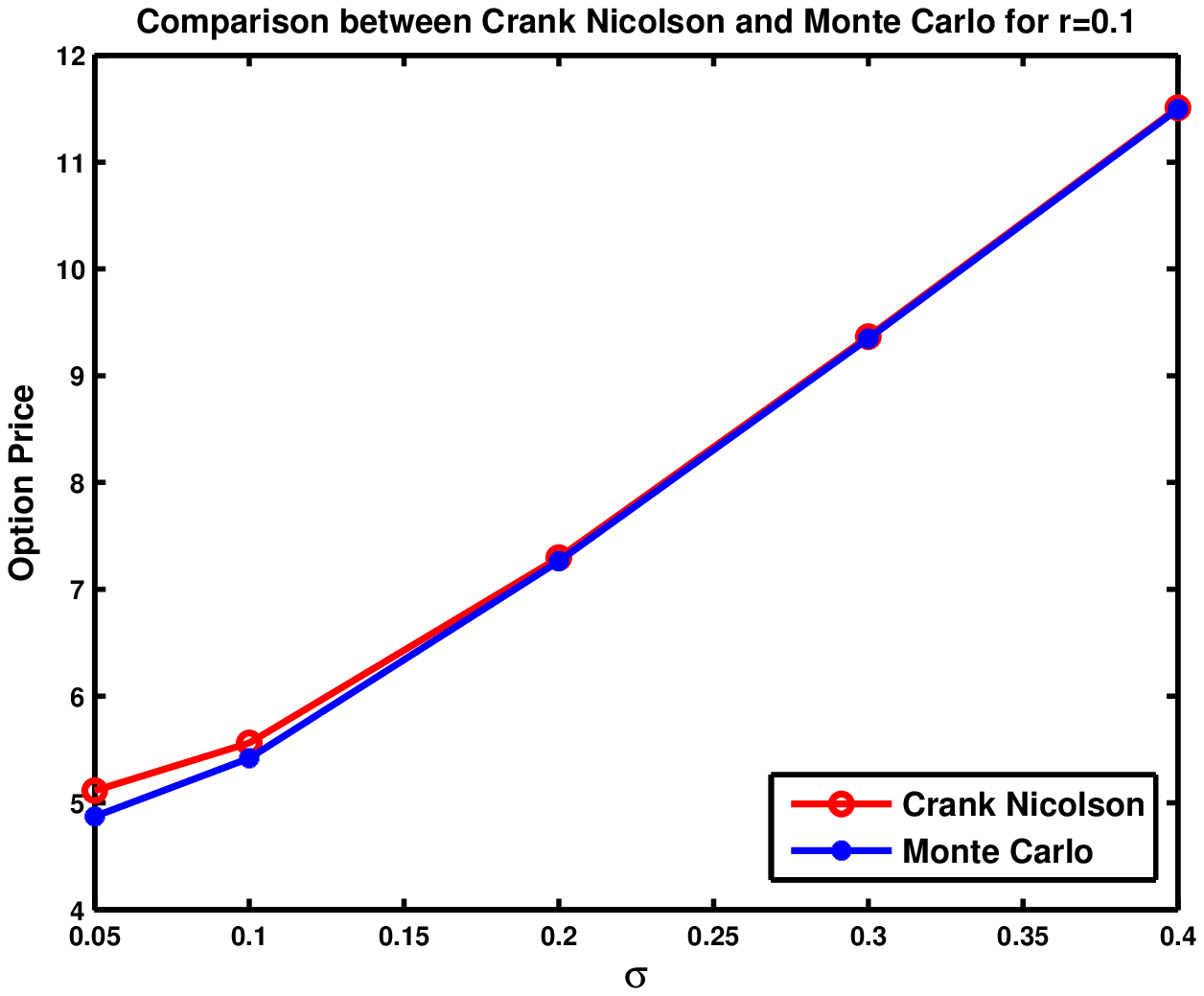}
\caption{Comparison between Monte Carlo Simulation (MC) and Crank Nicolson Implicit Method (CNIM)\label{MC_CNIM_Graph2} for r=0.1}
\end{figure}

\begin{figure}[hb]
\centering
\includegraphics[width=0.7\textwidth]{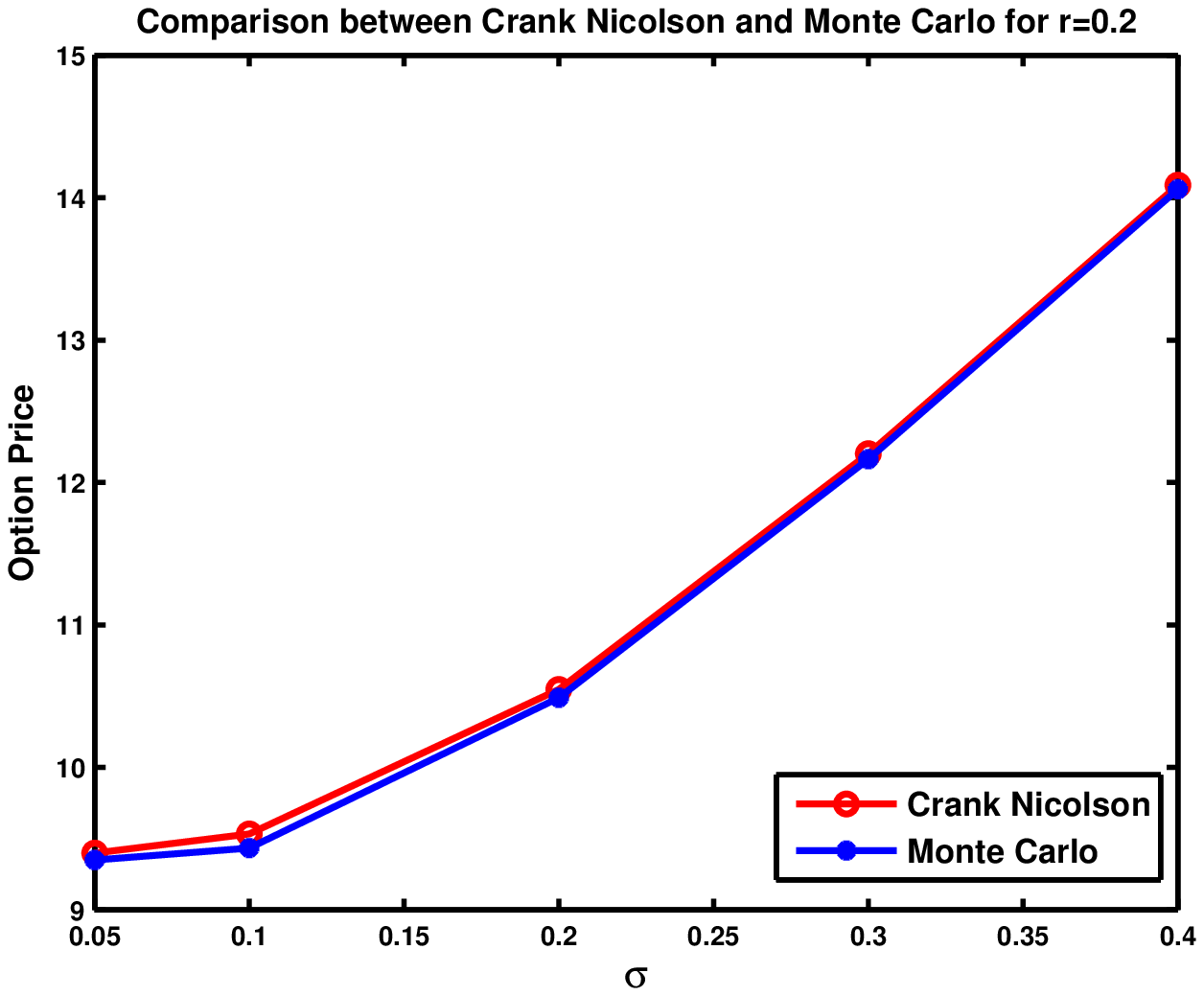}
\caption{Comparison between Monte Carlo Simulation (MC) and Crank Nicolson Implicit Method (CNIM)\label{MC_CNIM_Graph3} for r=0.2}
\end{figure}

\section{Higher Order Compact Scheme}

Finite difference methods have been used for solving ODE and PDE problems for quite a long time.
They are relatively easier to set up and solve, but require structured mesh \cite{Spotz94,Spotz95}. 
Finite element methods on the contrary are more sophisticated, works well with irregular domains and are amenable to unstructured meshes.
Finite element methods are relatively more challenging to implement.
Standard finite difference schemes like the CNIM (used in the previous section) are second order accurate. However,
in financial applications like option pricing a higher level of accuracy is desirable.
A direct extension of the central difference schemes to achieve higher order accuracy would involve more node points on
the stencil. An innovative way of achieving higher accuracy with lesser number of nodes in the stencil came by the way of
higher order compact (HOC) schemes. Spotz and Carey \cite{Spotz94} and Spotz \cite{Spotz95} provide an excellent discourse on 
application of this scheme in case of viscous flow and computational mechanics.

Despite it's enormous potential of application to finance problems, HOC schemes have not been used much in this area.
Zhao et al. \cite{Zhao07} presented a compact scheme for American option pricing with second order accuracy in space.
Tangman et al. \cite{Tangman08,Tangman08b} applied a HOC scheme for the pricing of American put option.
They do a comparative \cite{Tangman08} analysis with a non-compact fourth order scheme.
In their subsequent paper they describe an improvement of a method suggested by Han and Wu \cite{Han03}.
In this chapter we shall apply HOC scheme to the setup \eqref{eq:model} which is written in the following form \cite{BTP11},
\begin{equation}
c(R)\frac{\partial^2{H}}{\partial{R}^2}+d(R)\frac{\partial{H}}{\partial{R}}=g
\label{eq:neweqh}
\end{equation}
where $c(R)$ and $d(R)$ are as defined in the previous section and $g=-\frac{\partial H}{\partial t}$.
We now define notation $\delta$ \cite{Spotz94,Spotz95,Kalita02} as follows,
$$ \delta_R\:f := \frac{\partial f}{\partial R} \;,\qquad \delta^2_R\:f:= \frac{\partial^2 f }{\partial R^2} \quad \text{and so on.}$$
Thus rewriting Equation \eqref{eq:neweqh} in terms of finite difference discretization we get,
\begin{equation}
c_{i}\delta _{R}^{2}H_{i}+d_{i}\delta _{R}H_{i} = g_{i}
\label{eq:initpde}
\end{equation}
In the HOC scheme we derive the leading truncation error terms in terms of finite difference 
equation making use of the original equation. Denoting these terms by $\tau$, the HOC scheme is obtained by subtracting  
$\tau$ back to the original finite difference discretization. Thus we have,
\begin{equation}
c_{i}\delta _{R}^{2}H_{i}+d_{i}\delta _{R}H_{i}-\tau_i = g_{i}
\label{eq:newfde}
\end{equation}
where,
\begin{equation}
\tau_{i} = \frac{\Delta_{R}^{2}}{12}\left(c\frac{\partial^{4}H}{\partial R^{4}} + 2d\frac{\partial^{3}H}{\partial R^{3}}\right) + O(\Delta_{R}^{4})
\label{eq:truncation_error}
\end{equation}
From the initial PDE \eqref{eq:model} we get,
\begin{equation}
\frac{\partial c}{\partial R}\frac{\partial^{2}H }{\partial R^{2}} + c\frac{\partial^{3}H }{\partial R^{3}} + \frac{\partial d}{\partial R }\frac{\partial H}{\partial R} + d\frac{\partial^{2}H }{\partial R^{2}} = \frac{\partial g}{\partial R}
\label{eq:pde1}
\end{equation}
Therefore this can be rewritten in $\delta$ notation as,
\begin{equation}
\left. \frac{\partial^{3}H }{\partial R^{3}}\right|_{i} = -\frac{1}{c_{i}}\bigg[\delta_{R}c_{i}\delta_{R}^{2}H_{i} + \delta_{R}d_{i}\delta_{R}H_{i} + d_{i}\delta_{R}^{2}H_{i}\bigg] + \frac{1}{c_{i}}\delta_{R}g_{i}
\label{eq:te_part1}
\end{equation}
Differentiating the PDE \eqref{eq:pde1} w.r.t. $R$  once more and simplifying we get,
\begin{equation}
\frac{\partial^{2}c }{\partial R^{2}}\frac{\partial^{2}H }{\partial R^{2}} + 2\frac{\partial c}{\partial R}\frac{\partial^{3}H}{\partial R^{3} } + c\frac{\partial^{4}H }{\partial R^{4}} + \frac{\partial^{2}d }{\partial R^{2}}\frac{\partial H}{\partial R}
+ 2\frac{\partial d }{\partial R}\frac{\partial^{2}H}{\partial R^{2}} + d\frac{\partial^{3}H}{\partial R^{3}} = \frac{\partial^{2}g}{\partial R^{2}}
\label{eq:pde2}
\end{equation}
which can be written in $\delta$-notation as,
\begin{eqnarray}
-c_{i}\frac{\partial^{4}H }{\partial R^{4}}\biggr|_{i}&=&-\delta_{R}^{2}g_{i}+\delta_{R}^{2}c_{i}\delta_{R}^{2}H_{i}+\delta_{R}^{2}d_{i}\delta_{R}H_{i}+ 2\delta_{R}d_{i}\delta_{R}^{2}H_{i} \nonumber\\ 
&+&\left(\delta_{R}g_{i}-\delta_{R}c_{i}\delta_{R}^{2}H_{i}-\delta_{R}d_{i}\delta_{R}H_{i}-d_{i}\delta_{R}^{2}H_{i}\right)\frac{2\delta_{R}c_{i}}{c_{i}}
\label{eq:pde4}\\
&+&\left(\delta_{R}g_{i}-\delta_{R}c_{i}\delta_{R}^{2}H_{i}-\delta_{R}d_{i}\delta_{R}H_{i}-d_{i}\delta_{R}^{2}H_{i}\right)\frac{d_{i}}{c_{i}}\nonumber
\end{eqnarray}
Making use of approximations in equation (\ref{eq:te_part1}) and equation (\ref{eq:pde4}) in the truncation error ($\tau_{i}$)
(equation (\ref{eq:truncation_error})) and hence substituting $\tau_{i}$ in equation (\ref{eq:newfde}) we obtain,
\begin{eqnarray}
&&\delta_{R}^{2}H_{i}\left[ c_{i} + \frac{\Delta_{R}^2}{12}\frac{d_{i}^2}{c_{i}} + \sigma^{2}R_{i}\frac{\Delta_{R}^{2}d_{i}}{12c_{i}} + \frac{\Delta_{R}^2}{12}\sigma^{2} - \frac{r\Delta_{R}^{2}}{6} - \frac{\Delta_{R}^2}{6c_{i}}\sigma^{4}R_{i}^{2} - \frac{\Delta_{R}^2}{6c_{i}}d_{i}\sigma^{2}R_{i} \right] \nonumber\\
&&+\delta_{R}H_{i}\left[ d_{i} - \frac{r\Delta_{R}^{2}d_{i}}{12c_{i}} + \frac{r\Delta_{R}^{2}}{6}\frac{\sigma^{2}R_{i}}{c_{i}} \right]\\
&=&\left[ 1 + \frac{\Delta_{R}^2}{12}\delta_{R}^{2} + \frac{\Delta_{R}^2}{12}\frac{d_{i}}{c_{i}}\delta_{R} - \frac{\Delta_{R}^2}{6c_{i}}(\sigma^{2}R_{i})\delta_{R} \right]g_{i}\nonumber
\label{eq:HOC_part1_4}
\end{eqnarray}
Note that, $c_{i}=\frac{1}{2}\sigma^{2}R_{i}^{2}\Rightarrow \delta_{R}c_{i}=\sigma^{2}R_{i},\delta_{R}^{2}c_{i}=\sigma^{2}$ and
$d_{i}=1-rR_{i} \Rightarrow \delta_{R}d_{i}=-r,\delta_{R}^{2}d_{i}=0$.\\
We define,
\begin{eqnarray*}
A_{i}&=&\left[c_{i}+\frac{\Delta_{R}^2}{12}\frac{d_{i}^2}{c_{i}}+\sigma^{2}R_{i}\frac{\Delta_{R}^{2}d_{i}}{12c_{i}}+\frac{\Delta_{R}^2}{12}\sigma^{2}- \frac{r\Delta_{R}^{2}}{6} - \frac{\Delta_{R}^2}{6c_{i}}\sigma^{4}R_{i}^{2} - \frac{\Delta_{R}^2}{6c_{i}}d_{i}\sigma^{2}R_{i} \right]\\
B_{i}&=&\left[d_{i}-\frac{r\Delta_{R}^{2}d_{i}}{12c_{i}}+\frac{r\Delta_{R}^{2}}{6}\frac{\sigma^{2}R_{i}}{c_{i}}\right]\\
F1_{i}&=&\frac{\Delta_{R}}{24}\frac{d_{i}}{c_{i}}-\frac{\Delta_{R}}{12c_{i}}(\sigma^{2}R_{i})~,~k1=\frac{\Delta_{t}}{2\Delta_{R}^{2}}~,~
k2=\frac{\Delta_{t}}{4\Delta_{R}}.
\end{eqnarray*}
We now apply the HOC scheme to the above equation (recalling that $g=-\frac{\partial H}{\partial t}$) and obtain,
\begin{eqnarray}
&&k1\left( H_{i+1}^{n+1} - 2H_{i}^{n+1} + H_{i-1}^{n+1} + H_{i+1}^{n} - 2H_{i}^{n} + H_{i-1}^{n}\right)A_{i} \nonumber\\
&+& k2\left( H_{i+1}^{n+1} - H_{i-1}^{n+1} + H_{i+1}^{n} - H_{i-1}^{n} \right)B_{i}\nonumber\\
&=&\left( -H_{i}^{n+1} + H_{i}^{n} \right) - \frac{\Delta_{R}^{2}}{12}\left[ \frac{H_{i+1}^{n+1} - 2H_{i}^{n+1} + H_{i-1}^{n+1}}{\Delta_{R}^{2}} \right]\nonumber\\
&+&\frac{\Delta_{R}^{2}}{12}\left[ \frac{H_{i+1}^{n} - 2H_{i}^{n} + H_{i-1}^{n}}{\Delta_{R}^{2}} \right]\nonumber\\
&-&F1_{i}\left[H_{i+1}^{n+1} - H_{i-1}^{n+1}\right] + F1_{i}\left[H_{i+1}^{n}-H_{i-1}^{n}\right]
\label{eq:HOC_part2_1}
\end{eqnarray}
The above can now be rewritten as,
\begin{equation}
G_{i}H_{i+1}^{n} + K_{i}H_{i}^{n} + J_{i}H_{i-1}^{n}=D_{i}H_{i+1}^{n+1} + E_{i}H_{i}^{n+1} + F_{i}H_{i-1}^{n+1}
\label{eq:HOC_part2_5}
\end{equation}
where,
\begin{eqnarray*}
\begin{aligned}
G_{i} &= -k1A_{i} - k2B_{i} + \frac{1}{12} + F1_{i}&\\
K_{i} &= 2k1A_{i} + \frac{5}{6}&\\
J_{i} &= -k1A_{i} +k2B_{i} +  \frac{1}{12} - F1_{i}&
\end{aligned} \qquad \qquad
\begin{aligned}
D_{i} &= k1A_{i} + k2B_{i} + \frac{1}{12} + F1_{i}&\\
E_{i} &= -2k1A_{i} +\frac{5}{6}&\\
F_{i} &= k1A_{i} - k2B_{i}+\frac{1}{12} - F1_{i}.&
\end{aligned}
\end{eqnarray*}
The implicit method can be written in the matrix form,
\[BH^{(n)}=AH^{(n+1)} + b^{(n)},\]
where $H^{(n)}, A, B$ and $b^{(n)}$ has already been defined in the previous section.
As with the case of CNIM the number of time and space grid points were taken to be $101$ and $501$ respectively
along with the tolerance level of $\epsilon=10^{-8}$. The scheme was implemented using MatLab \TM.
\begin{table}[h!]
\begin{center}
\begin{tabular}{|l|l|l|l|l|l|l|}
\hline
\textbf{$r \rightarrow$} & $0.06$ & $0.06$ & $0.1$ & $0.1$ & $0.2$ & $0.2$ \\
\hline
\textbf{$\sigma\downarrow$} & \textbf{HOC} & \textbf{MC} & \textbf{HOC} & \textbf{MC} &
\textbf{HOC} & \textbf{MC} \\
\hline
$0.05$ & 3.1391	& 3.1509	& 4.8784	& 4.8734	& 9.3449	& 9.3486\\
& (0.393252)	& (4.511535)	& (0.385447)	& (4.492267)	& (0.367639)	& (4.556745)\\
\hline
$0.1$ & 3.8929	& 4.0124 &	5.3592 &	5.4183 &	9.4385 &	9.433\\
& (0.127529) &	(4.505087) &	(0.130137) &	(4.489719) &	(0.125133) &	(4.53052)\\
\hline
$0.2$ & 5.9919	& 6.1172	& 7.1641	& 7.2625	& 10.4486	& 10.4894\\
& (0.121434) &	(4.490006) &	(0.121898) &	(4.481982) &	(0.125619) &	(4.530701)\\
\hline
$0.3$ & 8.2462	& 8.3155	& 9.2902	& 9.3484	& 12.1361	& 12.163\\
& (0.123275) & (4.490814) &	(0.125051) & (4.518666) &	(0.125488) &	(4.500043)\\
\hline
$0.4$ & 10.4921	& 10.5358	& 11.4607	& 11.4952	& 14.0444 &	14.0581\\
& (0.12519) &	(4.483286) &	(0.120896) &	(4.483686) &	(0.120095) &	(4.490074)\\
\hline
\end{tabular}
\end{center}
\caption{Comparison between Monte Carlo Simulation (MC) and Higher Order Compact
Scheme (HOC). The values in braces represent the CPU time in seconds. The initial stock price was $S(0)=100$.}
\label{MC_HOC_Table}
\end{table}

\begin{figure}[hb]
\centering
\includegraphics[width=0.7\textwidth]{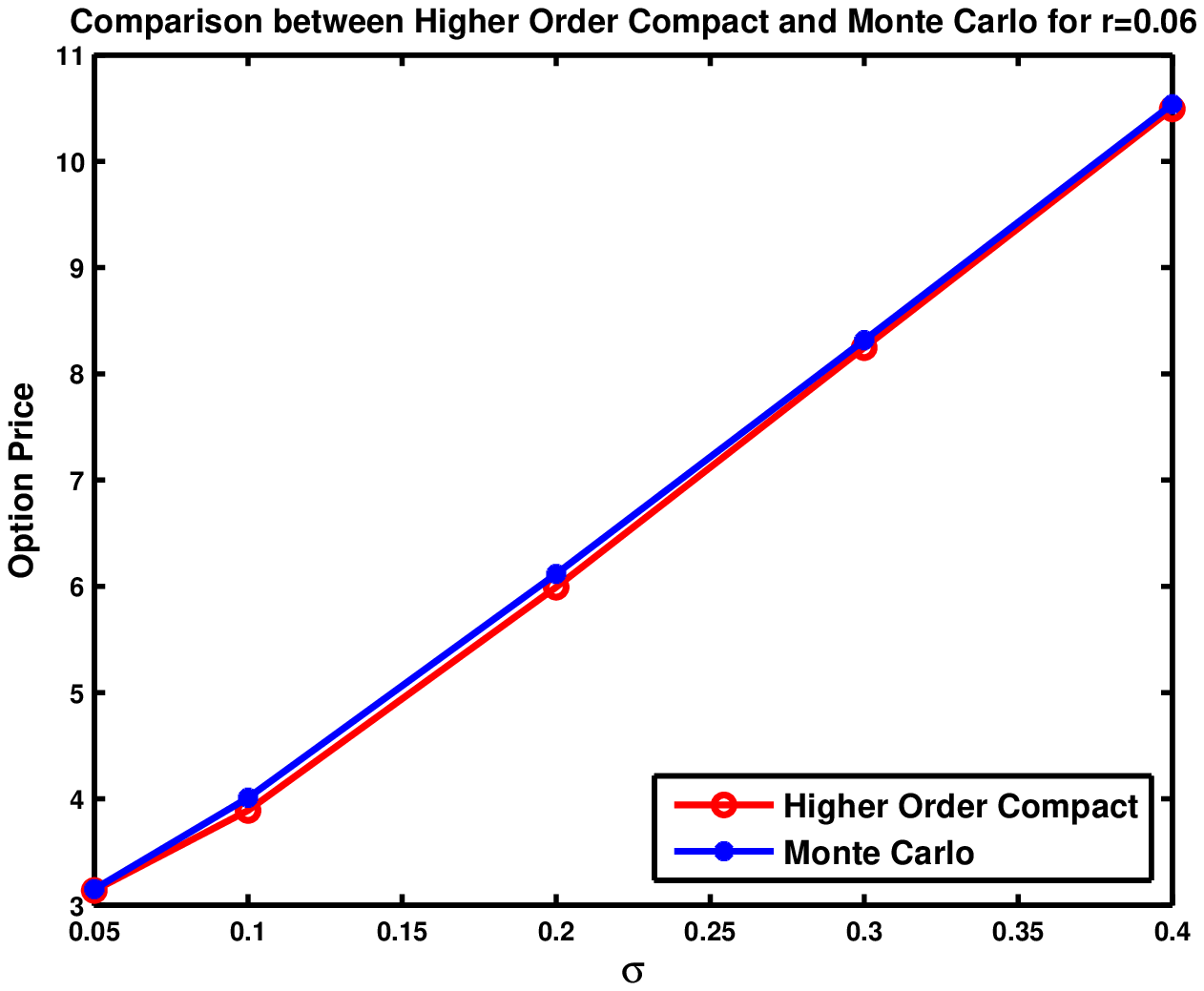}
\caption{Comparison between Monte Carlo Simulation (MC) and Higher Order Compact Scheme (HOC)\label{MC_HOC_Graph1} for r=0.06}
\end{figure}

\begin{figure}[hb]
\centering
\includegraphics[width=0.7\textwidth]{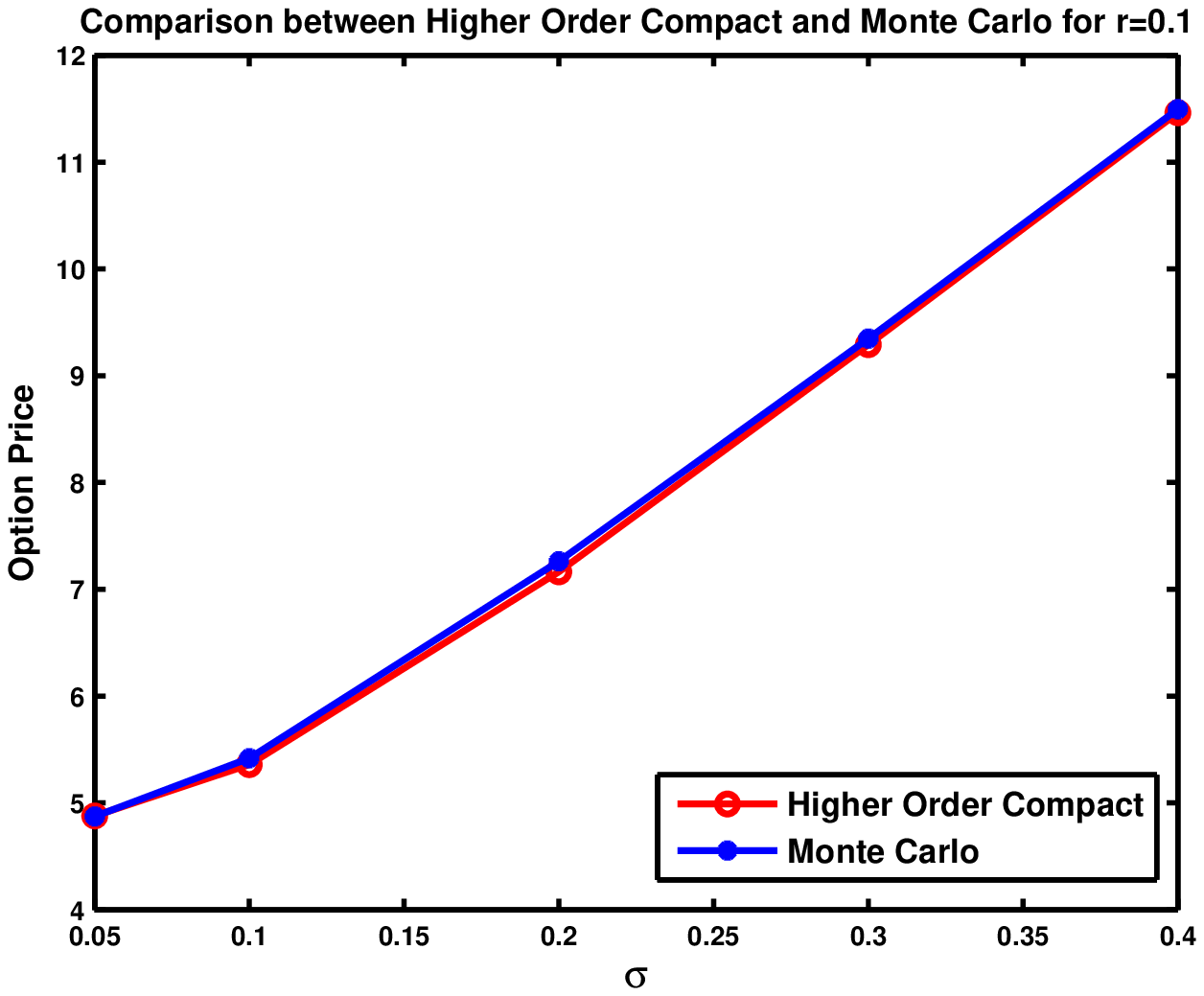}
\caption{Comparison between Monte Carlo Simulation (MC) and Higher Order Compact Scheme (HOC)\label{MC_HOC_Graph2} for r=0.1}
\end{figure}

\begin{figure}[hb]
\centering
\includegraphics[width=0.7\textwidth]{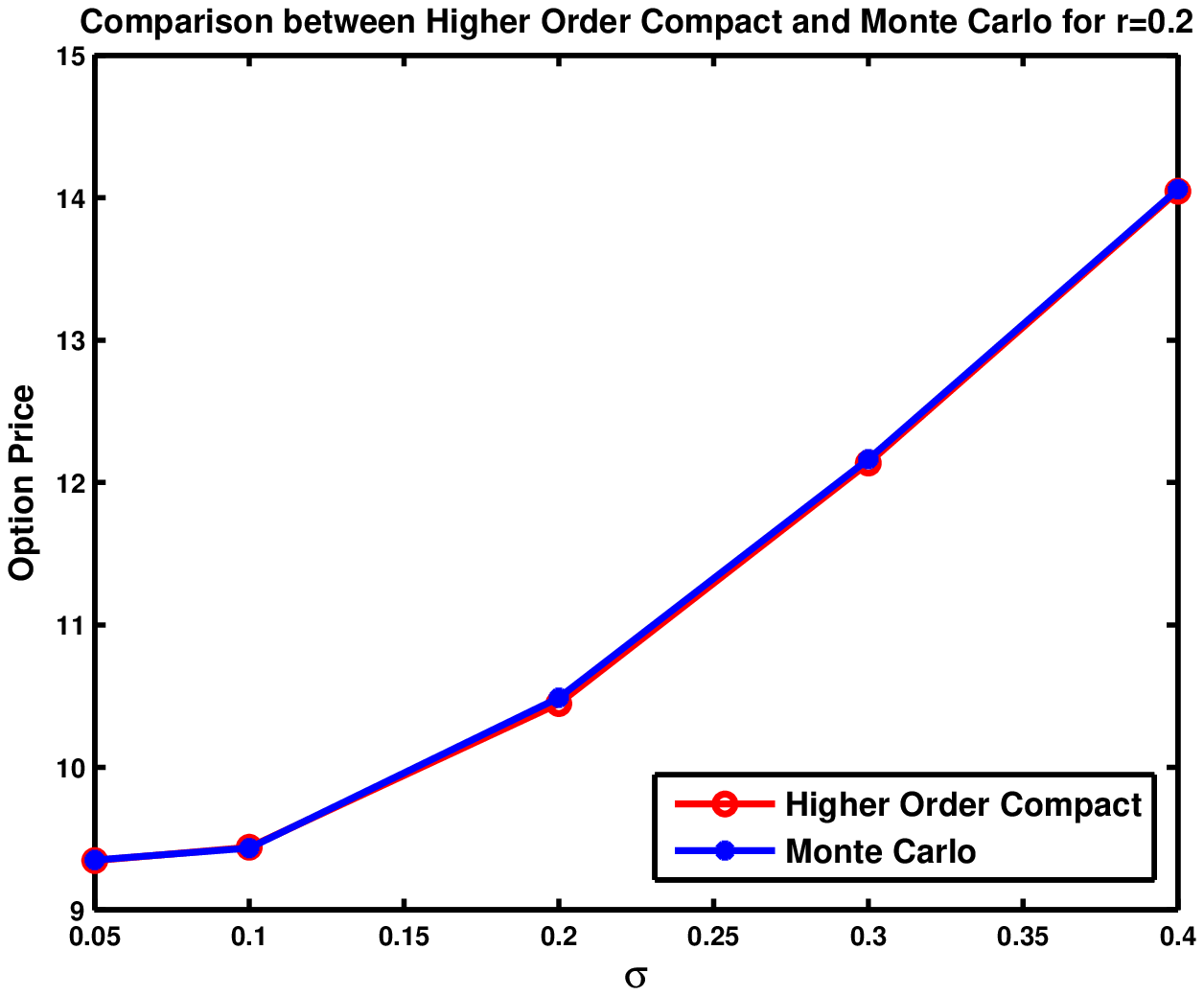}
\caption{Comparison between Monte Carlo Simulation (MC) and Higher Order Compact Scheme (HOC)\label{MC_HOC_Graph3} for r=0.2}
\end{figure}

\section{Results and Discussion}

In this section we discuss the results obtained by using the CNIM and the HOC schemes as outlined in the previous two sections.
As already noted we could not find any results for average strike Asian call option using numerical PDE methods.
For the purpose of comparison we used the Monte Carlo (MC) simulation as the benchmark value. 
We generated the path of a stock prices using the geometric Brownian motion process. We generated $50000$ such paths and determined
the option price from each of the paths generated. The average of all these option prices was taken to be the option price,
for the purpose of comparison with the
PDE methods. We generated the option prices using all the three methods for three values of
$r=0.06, 0.1, 0.2$ and five values of $\sigma=0.05,0.1,0.2,0.3,0.4$.

A comparative study of results from the CNIM and the MC methods showed a close match.
The comparative results are presented in Table
\ref{MC_CNIM_Table} along with the CPU time in seconds. The option prices for the three values of $r$ against the five values of
$\sigma$ have been presented in the graphical form in Figures (\ref{MC_CNIM_Graph1}), (\ref{MC_CNIM_Graph2}) and 
(\ref{MC_CNIM_Graph3}). For $r=0.06$ (Figure (\ref{MC_CNIM_Graph1})), the match was very close except the case where $\sigma=0.05$.
This slight difference in the option price is reduced when $r=0.1$ (Figure (\ref{MC_CNIM_Graph2})). The other values for $r=0.1$
showed a close match. The results were similar for the case $r=0.2$ except for a very minimal difference in the case of $\sigma=0.1$
(Figure (\ref{MC_CNIM_Graph3})).
The CPU time in case of CNIM was however considerably lower ($<0.5$ seconds) as compared with the Monte Carlo simulation ($>4$ seconds).

The results and comparison of the HOC scheme and the MC method indicates excellent consonance.
A comparison of the results from these two methods in terms of values and CPU time in seconds have been presented
in a tabular form in Table \ref{MC_HOC_Table}. The option prices from both the methods are very close to each other. In fact the
results obtained from the HOC scheme show a better match with the MC simulation results as compared with the CNIM method. This holds for 
all the values of $r$ and $\sigma$ and is evident from the comparative figures (Figures (\ref{MC_HOC_Graph1}), (\ref{MC_HOC_Graph2}), (\ref{MC_HOC_Graph3}))
of HOC and MC. As was the case with CNIM, the CPU time taken in case of the HOC scheme is significantly less ($<0.5$ seconds) 
in contrast to Monte Carlo simulation which requires  at least $4$ seconds.

\section{Conclusion}

In this paper we examined several ways of computing the price of an average strike Asian call option, namely Monte 
Carlo simulation and the numerical PDE approach.
In case of option pricing, the benchmark generally used is Monte Carlo simulation which suffers from some severe drawbacks like 
computational costs and a certain amount of uncertainty of pricing. In contrast, the usage of numerical PDE approaches 
that we have taken results in lesser computational costs and also provides an unique answer. The numerical PDE approach 
in pricing the average strike Asian call option is by and large an unexplored area, since this approach applied to 
Asian option is mostly concentrated on the case of fixed strike.
In this paper, we take the PDE approach to the pricing problem and present two schemes to accomplish this numerically.
Firstly, we use the Crank-Nicolson Implicit Method (which is second order) to solve the PDE and hence price the option.
Then, we present a Higher Order Compact scheme (fourth order) to solve the problem.
Finally we make a comparison of results obtained from the PDE approach with that of Monte Carlo.
The results obtained using the two PDE techniques were in excellent agreement with the Monte Carlo results.
The results obtained using Higher Order Scheme are closer to the Monte Carlo results as opposed to Crank-Nicolson Implicit method
vis-a-vis Monte Carlo. This is more so in case of lower values of $\sigma$.
To the best of our knowledge, this is the first work to use the numerical PDE approach for pricing Asian call options with average strike.
We believe this work would find more applications in the area of option pricing through the PDE approach.

\section{Acknowledgments}

The authors express their deep gratitude to Dr. Jiten C. Kalita (Department of Mathematics, IIT Guwahati) and 
Prof. Anoop K. Dass (Department of Mechanical Engineering, IIT Guwahati) for their valuable guidance and suggestions 
during the preparation of the manuscript.

\end{document}